\documentclass{article}
\usepackage[english]{babel}
\begin{document}
\title{An effective associative memory for pattern recognition}  
\author{B.V.Kryzhanovsky, L.B.Litinskii and A.Fonarev}
\date{ Institute of Optical Neural Technologies, Russian Academy of Sciences\\
44/2 Vavilov Street, 119333 Moscow\\
{\it kryzhanov@iont.ru, litin@iont.ru}\\
Department of Chemistry and Physics, Kean University,\\ 
Union, NJ 07083, USA}
\maketitle
\begin{abstract}
Neuron models of associative memory provide a new and prospective technology
for  
reliable date storage and patterns recognition. However, even when the patterns
are uncorrelated, the efficiency of most known models of associative memory is 
low. We developed a new version of associative memory with record 
characteristics of its storage capacity and noise immunity, which, in addition,
is effective when recognizing correlated patterns.
\end{abstract} 
\section{Introduction}
\label{sect:1}

Conventional neural networks can not efficiently recognize highly 
correlated
patterns. Moreover, they have small memory capacity. Particularly, 
the Hopfield 
neural network \cite{1} can store only $p_0\approx N/2\ln N$ 
randomized 
$N$-dimensional binary patterns. When the patterns are correlated 
this number  
falls abruptly. Few algorithms that can be used in this case 
(e.g., the
projection matrix method \cite{2}) are rather cumbersome and do 
not allow us 
to introduce a simple training principle \cite{3}. 
We offer a simple and effective algorithm that allows us to 
recognize a great
number  
of highly correlated binary patterns even under heavy noise 
conditions. We use
{\it a parametrical neural network} \cite{7}-\cite{12} that is a fully 
connected vector  
neural network similar to the Potts-glass network \cite{4} or 
optical network 
\cite{5},\cite{6}. The point of the approach is given below. 

Our goal is to 
recognize $p$ distorted $N$-dimensional {\it binary} patterns 
$\left\{ {Y^\mu  } \right\},\mu  \in \overline {1,p} $. 
The associative memory that meets this aim works in the following 
way. Each 
pattern $Y^\mu$ from the space ${\rm R}^{\rm N}$ is put in one-to-
one 
correspondence with an image $X^\mu$ from a space $\Re$ of greater 
dimensionality (Sect.~\ref{sect:3}). Then, the set of images 
$\left\{ {X^\mu  } \right\}$
is used to build a parametrical neural network 
(Sect.~\ref{sect:2}). 
The recognition
algorithm is as follows. An input binary vector $Y \in {\rm 
R}^{\rm N}$
is replaced by corresponding image $X \in \Re$, which is 
recognized by the 
parametrical neural network. Then  the result of recognition is  
mapped back into the original $N$-dimensional space. Thus, 
recognition of many 
correlated binary patterns is reduced to recognition of their 
images in the 
space $\Re$. Since parametrical neural network has an extremely  
large memory capacity (Sect.~\ref{sect:2}) and the mapping process 
$Y \to X$ allows us to 
eliminate the  correlations between patterns almost completely 
(Sect.~\ref{sect:3}), the problem is  
simplified significantly. 

\section{Parametrical neural network and their recognition 
efficiency}
\label{sect:2}

Let us described the parametrical 
neural network (PNN) \cite{7}-\cite{12}. We consider fully connected neural 
network of 
$n$ vector-neurons ({\it spins}). Each neuron is described by unit 
vectors 
$\vec x_i=x_i\vec e_{l_i }$, where an amplitude $x_i$ is equal 
$\pm 1$ 
and $\vec e_{l_i}$ is the $l_i$th unit vector-column of $Q$-dimensional space: 
$i = 1,...,n;\ 1\le l_i\le Q$. The state of the whole network is 
determined 
by the set of vector-columns $X = (\vec x_1 ,\vec x_2 ,...,\vec x_n )$. 
According \cite{11}, we define the Hamiltonian of the network the 
same as in the
Hopfield model 
$$H= -\frac12\sum\limits_{i,j = 1}^n {\vec x_i^+{\bf\hat 
T}_{ij}}\vec x_j ,
\eqno(1)$$
where $\vec x_i^+$ is the $Q$-dimensional vector-row. Unlike 
conventional 
Hopfield model, where the interconnections are scalars, in Eq.(1) 
an 
interconnection ${\bf \hat T}_{ij}$ is a $(Q\times Q)$-matrix 
constructed according  
to the generalized Hebb rule \cite{4}, 
$${\bf\hat T}_{ij}=(1-\delta_{ij})\sum\limits_{\mu=1}^p\vec 
x_i^\mu
\vec x_j^{\mu+},\eqno(2) $$
from $p$ initial patterns:
$$X^\mu=(\vec x_1^\mu,\vec x_2^\mu,...,\vec x_n^\mu),\ 
\mu=1,2,...,p.\eqno(3)$$
It is convenient to interpret the network with Hamiltonian (1) as 
a system of
interacting $Q$-dimensional spins and use the relevant 
terminology. In view of (1)-(3), the input signal arriving at the $i$th neuron 
(i.e., {\it 
the local field} that acts on $i$th spin) can be written as
$$\vec h_i=\sum\limits_{j = 1}^n {\bf\hat T}_{ij}\vec x_j=
\sum\limits_{l = 1}^Q A_l^{(i)} \vec e_l,\eqno(4)$$
where
$$A_l^{(i)}=\sum\limits_{j(\ne i)}^n{\sum\limits_{\mu=1}^p
{(\vec e_l\vec x_i^\mu)(\vec x_j^\mu\vec x_j )}},\ l = 
1,...,Q.\eqno(5)$$
The behavior of the system is defined in the natural way: under 
action of the
local field $\vec h_i$ the $i$th spin gets orientation that is as 
close to the 
local field direction as possible. In other words, the state of 
the $i$th 
neuron in the next time step, $t+1$, is determined
by the rule: 
$$\vec x_i (t + 1) = x_{\max } \vec e_{\max },\ x_{\max }={\rm 
sgn}
\left(A_{\max }^{(i)}\right),\eqno(6)$$
where {\it max} denotes the greatest in modulus amplitude 
$A_l^{(i)}$ in (5). 

The evolution of the network is a sequence of changes of neurons 
states according to (6) with energy (1) going down. Sooner or later the 
network finds itself at a fixed point. 

Let us see how efficiently PNN recognizes noisy
patterns. Let the distorted $m$th pattern $X^m$ come to the system input, 
i.e. the neurons are in the initial 
states described as $\vec x_i  = \hat a_i\hat b_i\vec x_i^m$. Here 
$\hat a_i$
is the multiplicative noise operator, which changes the sign of 
the amplitude
$x_i^m$ of the vector-neuron $\vec x_i^m  = x_i^m \vec e_{l_i^m }$ with 
probability 
$a$ and keeps it the same with probability $1-a$; the operator 
$\hat b_i$
changes the basis vector $\vec e_{l_i^m}\in\{\vec e_l\}_1^Q$
by any other from $\{\vec e_l\}_1^Q$ with probability $b$ and 
retains it 
unchanged with probability $1-b$. 
In other words, $a$ is the probability of an error in the sign of 
the neuron
("-" in place "+" and vice versa), $b$ is the probability of an 
error in the 
vector state of the neuron.
The network recognizes the reference pattern 
$X^m$ correctly, if the output of the $i$th neuron defined by 
Eq.(6) is equal 
to $\vec x_i^m$, that is $\vec x_i=\vec x_i^m$. Otherwise, PNN 
fails 
to recognize the pattern $X^m$. According to the Chebyshev-Chernov 
method 
\cite{9} (for such problems it is described comprehensively in 
\cite{7}-\cite{12}, \cite{6}) the probability of recognition failure is
$${\rm P}_{err}\le n\exp \left[ { - \frac{{nQ^2 }}{{2p}}(1 -
2a)^2 (1 -  
b)^2 } \right].\eqno(7)$$
The inequality sets the upper limit for the probability of 
recognition failure
for PNN. The memory capacity of PNN 
(i.e., the greatest number of patterns that can be recognized) is 
found from 
(7):
$$p_c  = nQ^2 \frac{{(1 - 2a)^2 (1 - b)^2 }}{{2\ln n}}.\eqno(8)$$

Comparison of (8) with similar expressions for the Potts-glass neural network 
\cite{4} and the 
optical network  \cite{5},\cite{6} shows that the memory capacity 
of PNN is 
approximately twice as large as the memories of both 
aforementioned models. 
That means that under other conditions being equal its recognition 
power is 
20-30\% higher.

It is seen from (7) that with growing $Q$ the noise immunity of 
PNN  
increases exponentially. The memory capacity also grows: it is 
$Q^2$ times as large as that of the Hopfield network. For example, 
if $Q=32$, 
the 180-neuron PNN can recognize 360 patterns ($p/n = 2$) with 
85\% 
noise-distorted components (Fig.1). With smaller 
distortions,  
$b=0.65$, the same network can recognize as many as 1800 patterns 
($p/n=10$), 
{\it etc}. 
Let us note, that some time ago the restoration of 30\% noisy 
pattern
for $p/n = 0.1$  was a demonstration of the best recognition 
ability of the 
Hopfield model \cite{10}.  

Of course, with regard to calculations, PNN is more complex than the 
Hopfield model. On the other hand the computer code can be done rather 
economical with the aid of extracting bits only. It is not necessary 
to keep a great number of $(Q\times Q)$-matrices ${\bf \hat T}_{ij}$ 
(2) in your computer memory. Because of the complex structure of neurons, 
PNN works $Q$-times slower than the Hopfield model, but it makes possible 
to store $Q^2$-times greater number of patterns. In addition, the Potts-glass 
network operates $Q$-times slower than PNN. 
 
Fig.2 shows the recognition reliability 
$\bar P=1-{\rm P}_{err}$ as a 
function of the noise intensity $b$ when the number of patterns is 
twice the 
number of neurons ($p=2n$) for $Q=8, 16, 32$. We see that when the noise 
intensity is less 
than a threshold value 
$$b_c=1-\frac2Q\sqrt{\frac{p}{n}},$$
the network demonstrates reliable recognition of noisy patterns. 
We would like to use these outstanding properties of PNN for 
recognition of
correlated binary patterns. The point of our approach is given in 
next Sections.

\section{Mapping algorithm}
\label{sect:3}

Let $Y = (y_1 ,y_2 ,...,y_N )$ be an $N$-dimensional binary 
vector, 
$y_i  = \left\{\pm 1 \right\}$. We divide it mentally into $n$ 
fragments of $k+1$ elements each, $N = n(k + 1)$. 
With each fragment we associate an integer number $\pm l$ according the rule:
the first element of the fragment defines the sign of the number, 
and the other $k$ elements determine the absolute value of 
$l$:
$$l=1+\sum_{i=1}^k(y_i+1)\cdot 2^{i-2}; \quad 1\le l\le 2^k.$$ 
Now we associate each fragment with a vector $\vec x =  \pm \vec e_l$, where 
$\vec e_l$ is the $l$th unit vector of the real space ${\rm 
R}^{\rm Q}$ 
and $Q = 2^k$. We see that any binary vector $Y \in {\rm R}^{\rm 
N}$
one-to-one corresponds to a set of $n$ $Q$-dimensional unit 
vectors, 
$X = (\vec x_1 ,\vec x_2 ,...,\vec x_n)$, which we call {\it the 
internal 
image} of the binary vector $Y$. (In the next Section we 
use the internal images $X$ for PNN construction.) The number $k$ is called 
{\it a mapping parameter}. 

For example, the binary vector  
$Y=(-1,1,-1,-1,1,-1,-1,1)$ can be split into two fragments of four elements: 
(-11-1-1) and
(1-1-11); the mapping parameter $k$ is equal 3, $k=3$. 
The first fragment ("-2" in our notification) corresponds to 
the 
vector $-\vec e_2$ from the space of dimensionality 
$Q=2^3=8$, and 
the second fragment ("+5" in our notification) corresponds to 
the vector 
$+ \vec e_5  \in {\rm R}^{\rm 8}$. The relevant  
mapping can be 
written as $Y \to X = (-\vec e_2 ,+\vec e_5)$. 

It is important that the mapping is biunique, i.e., the binary 
vector $Y$ can be
restored uniquely from its internal image $X$. It is even more 
important that the 
mapping eliminates correlations between the original binary 
vectors. For 
example, suppose we have two 75\% overlapping binary vectors 
$$\begin{array}{rrrrrrrrrl}Y_1=(&1,&-1,&-1,&-1,&-1,&-1,&-1,&1&)\\
Y_2=(&1,&-1,&-1,&1,&1,&-1,&-1,&1&)\end{array}.$$ 
Let us divide each vector into four 
fragments of two 
elements. In other words, we map these vectors with the mapping 
parameter 
$k = 1$. As a result we obtain two internal images 
$X_1=(+\vec e_1,-\vec e_1,-\vec e_1,-\vec e_2)$ and 
$X_2=(+\vec e_1,-\vec e_2,+\vec e_1,-\vec e_2)$ with $\vec e_l\in{\rm R^2}$. 
The overlapping of these 
images is 50\%. If the mapping parameter $k = 3$ is used, the 
relevant images 
$X_1  = ( + \vec e_1 , - \vec e_5 )$ and 
$X_2  = ( + \vec e_5 , + \vec e_5 )$ with $\vec e_l\in{\rm R^8}$ 
do not overlap at all.

\section{Recognition of the correlated binary patterns}
\label{sect:4}

In this Section we describe the work of our model as a whole, i.e. 
the mapping
of original  
binary patterns into internal images and recognition of these 
images with the 
aid of PNN. For a
given mapping  
parameter $k$ we apply the procedure from Sect.~\ref{sect:3} to a 
set of 
binary patterns
$\left\{{Y^\mu}\right\}\in{\rm R}^{\rm N},\ \mu\in\overline 
{1,p}$. As a result 
we obtain a set of internal images $\left\{{X^\mu}\right\}\in 
\Re$, 
allowing us to build PNN with $n = N/(k + 1)$ 
$Q$-dimensional vector-neurons, where $Q = 2^k$. Note, the 
dimension $Q$ of vector-neurons increases exponentially with $k$ 
increasing, and this improves the properties of PNN. 

In further analysis we use 
so-called {\it biased} patterns $Y^\mu$ whose components $y_i^\mu$
are either $+1$ or $-1$ with probabilities $(1 + \alpha )/2$ and 
$(1 - \alpha )/2$ respectively $(-1\le\alpha\le 1)$. 
The patterns will have a mean activation of $\overline{Y^\mu}=\alpha$
and a mean correlation $\overline{(Y^\mu,Y^{\mu\prime})}=\alpha^2,\ 
\mu\ne\mu\prime$. 
Our aim is to recognize 
the $m$th noise-distorted binary pattern 
$\tilde Y^m=(s_1y_1^m,s_2y_2^m,...,s_Ny_N^m)$, where a random 
quantity 
$s_i$ changes the variable $y_i^m$ with probability $s$, and the 
probability 
for $y_i^m$ to remain the same is $1-s$. In other words, $s$ is  
the distortion level of the $m$th binary pattern. Mapped into space 
$\Re$, this binary 
vector turns into the $m$th noise-distorted internal image $\tilde 
X^m$
, which has to be recognized by PNN. Expressing
multiplicative noises $a$ and $b$ as functions of $s$ and 
substituting the result in (7), we find that the  
probability of misrecognition of the internal image $X^m$ is
$${\rm P}_{err}=n\left(\mbox{ch}(\nu\alpha^3)-
\alpha\mbox{sh}(\nu\alpha^3)
\right)  
\cdot\exp\left[-\frac{\nu}{2\beta}(1+\beta^2\alpha^6)\right], 
\eqno(9)
$$
where
$$\nu=n(1 - 2s)^2(1 - s)^k,\ \beta=p[A/(1 - s)]^k,\ 
A = (1+\alpha^2)[1+\alpha^2(1-2s)]/4.$$

Expression (9) describes the Hopfield model when $k=0$. In this 
case, even
without a bias ($\alpha=0$)  
the memory capacity does not exceed a small value $p_0\approx 
N/2\ln N$. However, even if small correlations ($\alpha>N^{-1/3}$) are present, the number of 
recognizable patterns is $(N\alpha^3)$-times reduced. In 
other words, 
the network almost fails to work as associative memory. The 
situation  
changes noticeably when parameter $k$ grows: the memory capacity 
increases and 
the influence of correlations decreases. In particular, when 
correlations are 
small ($\nu\alpha^3<1$), we estimate memory capacity from (9) as
$$p=p_0\frac{[(1 - s)^2 /A]^k}{k+1}.$$
Let us return to the example from \cite{10} with $\alpha=0$. When $k=5$, 
in the framework of our approach one can use 5-times greater number 
of randomized binary patterns, and when $k=10$, the number of the patterns 
is 80-times greater.  

When the degree of correlations is high ($\nu\alpha^3>1$), the 
memory capacity is somewhat smaller:
$$p=\alpha^{-3}[(1-s)/A]^k.$$
Nevertheless, in either case increasing $k$ gives us {\it the 
exponential} growth of the number of  
recognizable patterns and a rise of recognition reliability. 

Fig.3 shows the growth of the  
number of recognizable patterns when $k$ increases for distortions 
$s$ of 10\% to 50\% ($\alpha=0$). 

Fig.4 demonstrates how the destructive 
effect of
correlations diminishes when $k$ increases: the recognition power 
of the 
network is zero when $k$ is
small; however, when the mapping parameter $k$ reaches a certain 
critical 
value, the misrecognition probability falls sharply (the curves 
are drawn for 
$\alpha=0, 0.2, 0.5, 0.6$, $p/N=2$ and $s=30\%$).
Our computer simulations confirm these results.

\section{Concluding remarks}
\label{sect:5}

Our algorithm using the large memory capacity of PNN and its high 
noise 
immunity, allows us to recognize many correlated 
patterns reliably. The method also works when all patterns have 
the same 
fragment or fragments. For instance, for $p/N=1$ and $k=10$, this 
kind of 
neural network permits reliable recognition of patterns
with 40\% coincidence of binary components. 

When initially there is a set of 
correlated vector-neuron patterns which must be stored and 
recognized, the 
following algorithm can be used to suppress correlations. At 
first, the 
patterns with vector coordinates of original dimensionality $q_1$ 
are 
transformed into binary patterns. Then they are mapped into 
patterns with
vector neurons of dimensionality $q_2 (q_2 > q_1)$. A PNN is built 
using patterns with $q_2$-dimensional vector neurons. 
This double mapping eliminates correlations and provides a  
reliable pattern recognition. 

In conclusion we would like to point out that a reliable  
recognition of correlated patterns requires a random numbering of 
their
coordinates (naturally, in the same fashion for all the patterns). 
Then the 
constituent vectors do not have large identical fragments. 
In this case a decorrelation of the patterns can be done with  
a relatively small mapping parameter $k$ and, therefore, takes 
less 
computation time. 

The work was supported by Russian Basic Research Foundation 
(grant 01-01-00090), the program "Intellectual Computer Systems" (the 
project 2.45) and by President of Russian Federation (grant SS-1152-2003-1).

\vskip 10mm

\centerline{\bf List of captions:}
\vskip 5mm
Fig.1. Process of recognition of letter "A" by PNN at $p/n=2, Q=32, b=0.85$.
Noise-distorted pixels are marked in gray.

Fig.2. Recognition reliability of PNN $\bar P=1-{\rm P}_{err}$ 
as a function of noise intensity $b$.

Fig.3. Growth of memory capacity with the mapping parameter $k$ 
($s=0.1 - 0.5$).

Fig.4. Fall of the misrecognition probability with the mapping parameter 
$k$.

\begin{thebibliography}{99}

\bibitem{1} Hopfield, J.J.: Neural networks and physical systems 
with emergent
collective computational abilities. Proc. Natl. Acad. Sci. {\it 
USA} {\bf 79} 
(1982) 2554-2558.

\bibitem{2} Personnaz, L., Guyon, I., Dreyfus, G.: 
Collective computational properties of neural networks: New 
learning 
mechanisms. Phys. Rev. A {\bf 34} (1986) 4217-4228. 

\bibitem{3} Mikaelian, A.L., Kiselev, B.S., Kulakov, N.Yu., 
Shkitin, V.A. and 
Ivanov, V.A.: Optical implementation of high-order associative 
memory.
Int. J. of Opt. Computing {\bf 1} (1990) 89-92.

\bibitem{7} Kryzhanovsky, B.V., Litinskii, L.B., Fonarev A.:
Optical neural network based on the parametrical four-wave mixing 
process.
In: Wang, L., Rajapakse, J.C, Fukushima, K., Lee, S.-Y., and Yao, 
X. (eds.):
Proceedings of the 9th International Conference on Neural 
Information
Processing (ICONIP'02), Orchid Country Club, Singapore, (2002) 
{\bf 4} 1704-1707.

\bibitem{11} Kryzhanovsky, B.V., Litinskii, L.B., Mikaelyan A.L.
Parametrical neural network. Cond-mat/0212541.

\bibitem{12} Kryzhanovsky, B.V., Litinskii, L.B., Mikaelyan A.L.
Parametrical neural network. Optical Memory \& Neural Networks, {\bf 12}, \# 3 
(in press).

\bibitem{4} Kanter, I.: Potts-glass models of neural networks. 
Phys. Rev. A {\bf 37} (1988) 2739-2742.

\bibitem{5} Kryzhanovsky, B.V., Mikaelian, A.L.: 
On the Recognition Ability of a Neural Network on
Neurons with Parametric Transformation of Frequencies.
Doklady Mathematics, {\bf 65(2)} (2002) 286-288.

\bibitem{6} Kryzhanovsky, B.V., Kryzhanovsky, V.M., Mikaelian, 
A.L., 
Fonarev, A.: Parametric dynamic neural network recognition power.
Optical Memoryu \& Neural Networks {\bf 10} (2001) 211-218.

%\bibitem{8} Hebb, D.O.: The Organization of Behavior. Wiley, New York (1949)

\bibitem{9} Chernov, N.: A measure of asymptotic efficiency for 
tests of 
hypothesis based on the sum of observations. Ann. Math. Statistics 
{\bf 23} 
(1952) 493-507.

\bibitem{10} Kinzel, W.: Spin Glasses and Memory.
Physica Scripta {\bf 35} (1987) 398-401.
\end{thebibliography}
\end{document}